\documentclass{article}\usepackage[]{graphicx}\usepackage[]{color}
\makeatletter
\def\maxwidth{ %
  \ifdim\Gin@nat@width>\linewidth
    \linewidth
  \else
    \Gin@nat@width
  \fi
}
\makeatother

\definecolor{fgcolor}{rgb}{0.345, 0.345, 0.345}

\usepackage{framed}
\makeatletter
 {\par\unskip\endMakeFramed%
 \at@end@of@kframe}
\makeatother

\definecolor{shadecolor}{rgb}{.97, .97, .97}
\definecolor{messagecolor}{rgb}{0, 0, 0}
\definecolor{warningcolor}{rgb}{1, 0, 1}
\definecolor{errorcolor}{rgb}{1, 0, 0}
\newenvironment{knitrout}{}{} 

\usepackage{alltt} 

\usepackage{etoolbox}
\newtoggle{arxivformat}
\toggletrue{arxivformat}

\iftoggle{arxivformat} {
  \usepackage[sort&compress, numbers]{natbib}
  \usepackage{times}
}{
  \usepackage{nips15submit_e,times}
  \usepackage[sort&compress, numbers]{natbib}
}

\usepackage{hyperref}
\usepackage{url}

\usepackage[margin=1.5in]{geometry}

\usepackage{times}
\usepackage{graphicx} 
\usepackage{subcaption}
\usepackage{caption}

\usepackage{amsmath}
\usepackage{amsthm}
\usepackage{appendix}

\usepackage{amssymb}
\usepackage{bbm}

\newcommand{\app}[1]{Appendix~\ref{app:#1}}

\newcommand{\mysec}[1]{Section~\ref{sec:#1}}

\newcommand{\eq}[1]{Eq.~(\ref{eq:#1})}

\newcommand{\fig}[1]{Fig.~(\ref{fig:#1})}

\newcommand{\constant}{C} 

\newcommand{\mbeq}{\mathbb{E}_{q}}

\newcommand{\cov}{\textrm{Cov}}

\newcommand{\pthetapost}[1][\alpha]{p_{x}^{#1}}
\newcommand{\qthetapost}[1][\alpha]{q_{x}^{#1}}
\newcommand{\gtheta}{g\left(\theta\right)}
\newcommand{\expectp}[2][\alpha]{\mathbb{E}_{\pthetapost[#1]}\left[#2\right]}
\newcommand{\expectq}[2][\alpha]{\mathbb{E}_{\qthetapost[#1]}\left[#2\right]}
\newcommand{\epgtheta}[1][\alpha]{\expectp[#1]{\gtheta}}

\theoremstyle{plain}

\title{Robust Inference with Variational Bayes}

\author{
Ryan Giordano\\
Department of Statistics\\
University of California, Berkeley\\
Berkeley, CA 94720 \\
\texttt{rgiordano@berkeley.edu}
\and
Tamara Broderick \\
Department of EECS\\
Massachusetts Institute of Technology\\
Cambridge, MA 02139\\
\texttt{tbroderick@csail.mit.edu}
\and
Michael Jordan \\
Department of EECS\\
University of California, Berkeley\\
Berkeley, CA 94720 \\
\texttt{jordan@cs.berkeley.edu }
}

\iftoggle{arxivformat}
{}{
  \nipsfinalcopy 
}
\IfFileExists{upquote.sty}{\usepackage{upquote}}{}
\begin{document}

\maketitle

\section{Introduction}\label{sec:intro}

In Bayesian analysis, the posterior follows from the data and a choice of a
prior and a likelihood.  One hopes that the posterior is robust to reasonable
variation in the choice of prior and likelihood, since this choice is made by
the modeler and is necessarily somewhat subjective. For example, the process of
prior elicitation may be prohibitively time-consuming,  two practitioners may
have irreconcileable subjective prior beliefs,  or the model may be so complex
and high-dimensional that humans cannot reasonably express their prior beliefs
as formal distributions.  All of these circumstances might give rise to a range
of reasonable prior choices. If the posterior changes substantially with these
choices of prior, then the analysis lacks objectivity. Measuring the sensitivity
of the posterior to variation in the likelihood and prior is the central concern
of the field of \emph{robust Bayes}. A robust posterior is one that does not
depend strongly on reasonable variation in the choice of model or prior, and
robust Bayes provides methods for quantifying posterior robustness
\citep{insua:2012:robust}.

Despite the fundamental importance of the problem and a considerable body of
literature, the tools of robust Bayes are not commonly used in practice. This is
in large part due to the difficulty of calculating robustness measures from MCMC
draws\citep{berger:2012:robust, roos:2015:sensitivity}. Although methods for
computing robustness measures from MCMC draws exist, they lack generality and
often require additional coding or computation \footnote{See \app{mcmc} for a
literature review.}. Consequently, formal robust Bayes methods are least used in
complex, hierarchical models, exactly when they are needed most. Instead,
modelers are tempted to either compute ad-hoc robustness estimates (e.g. by
manually changing the priors and re-running their chain) or to ignore the
problem altogether.

In contrast to MCMC, variational Bayes (VB) techniques are readily amenable to
robustness analysis.  The derivative of a posterior expectation with respect to
a prior or data perturbation is a measure of \emph{local robustness} to the
prior or likelihood \citep{gustafson:2012:localrobustnessbook}. Because VB casts
posterior inference as an optimization problem, its methodology is built on the
ability to calculate derivatives of posterior quantities with respect to model
parameters, even in very complex models. Variational methods for posterior
approximation are increasingly providing a scalable alternative to MCMC for
posterior approximation, and this offers the opportunity to bring fast,
easy-to-use robustness measures into common practice.

In the present work, we develop local prior robustness measures for
\emph{mean-field variational Bayes} (MFVB), a VB technique which imposes a
particular factorization assumption on the variational posterior approximation.
In past work \citep{giordano:2015:lrvb}, we demonstrated that a MFVB analysis can
be quickly and straightforwardly augmented to provide information about local
perturbations of the posterior variational approximation using linear response
methods from statistical physics.  We show that this framework can be extended
to provide fast, easy-to-use prior robustness measures for posterior inference
and thereby bring robustness analysis into common Bayesian practice.

In the remainder of the present work, we start by outlining existing local prior
measures of robustness in \mysec{robustness_measures}. We extend the linear
response techniques of \citep{giordano:2015:lrvb} in \mysec{lrvb_formulas}.  In
\mysec{lrvb_robustness} we use these results to derive closed-form measures of
the sensitivity of mean-field variational posterior approximation to prior
specification. In \mysec{experiments} we demonstrate our method on a
meta-analysis of randomized controlled interventions in access to microcredit in
developing countries.

\section{Robustness measures}\label{sec:robustness_measures}

Denote our $N$ data points by $x = (x_1, \ldots, x_N)$ with $x_n \in
\mathbb{R}^{D}$. Denote our parameter by the vector $\theta \in \mathbb{R}^{K}$.
We denote the prior parameters by $\alpha$, where either $\alpha \in
\mathbb{R}^{M}$ or $\alpha$ may be function-valued. Let $\pthetapost$ denote the
posterior distribution of $\theta$, as given by Bayes' Theorem:
\begin{eqnarray*}
  \pthetapost\left(\theta\right) :=
  p\left(\theta \vert x, \alpha \right) =
    \frac{p\left(x \vert \theta \right) p\left(\theta \vert \alpha \right)}
    {p\left(x\right)}.
\end{eqnarray*}
A typical end product of a Bayesian analysis might be a posterior expectation of
some function $\gtheta$ (e.g., a mean or variance): $\epgtheta$, which is a
functional of $g$. We suppose that we have determined that the prior parameter
$\alpha$ belongs to some set $\mathcal{A}$, perhaps after expert prior
elicitation. Finding the extrema of $\epgtheta$ as $\alpha$ ranges over all of
$\mathcal{A}$ is intractable or difficult except in special cases
\citep{moreno:2012:globalrobustness}. An alternative is to examine how much
$\epgtheta$ changes locally in response to small perturbations in the value of
$\alpha$:
\begin{eqnarray}\label{eq:local_robustness}
\left. \frac{d\epgtheta}{d\alpha} \right|_{\alpha} \Delta \alpha
\end{eqnarray}
That is, we consider \emph{local robustness}
\citep{gustafson:2012:localrobustnessbook} properties in lieu of global ones.
When $\alpha$ is function-valued, we take \eq{local_robustness} to be a Gateaux
derivative.  By calculating \eq{local_robustness} for all
$\Delta \alpha \in \mathcal{A} - \alpha$, we can estimate the robustness
of $\epgtheta$ in a small neighborhood of $\alpha$.

\section{Linear response variational Bayes and extensions}
\label{sec:lrvb_formulas}

We next review and extend linear response perturbations to a mean-field
variational Bayes posterior approximation \citep{giordano:2015:lrvb} in order to
quickly and easily evaluate \eq{local_robustness}. Let $\qthetapost$ denote the
variational approximation to posterior $\pthetapost$. Recall that $\qthetapost$
is an approximate distribution selected to minimize the Kullback-Liebler
divergence between $\pthetapost$ and $q$ across distributions $q$ in some class
$\mathcal{Q}$. We consider the case where the variational family, $\mathcal{Q}$,
is a class of products of exponential family distributions
\citep{bishop:2006:pattern}:
\begin{eqnarray}\label{eq:kl_minimization}
\qthetapost &:=&
  \textrm{argmin}_{q \in \mathcal{Q}} \left\{S - L\right\}
  \quad \textrm{for} \quad
\mathcal{Q} = \left\{q: q(\theta)  = \prod_{k=1}^K q(\theta_k); \quad \forall k,
  q(\theta_k) \propto \exp(\eta_k ^T \theta_k) \right\}
\nonumber\\
  L &:=&   \mbeq\left[ \log p\left(x \vert \theta \right)\right] +
           \mbeq\left[ \log p\left(\theta \vert \alpha \right) \right]
,\; \quad
  S :=   \mbeq\left[ \log q \left(\theta \right) \right]
\end{eqnarray}
We assume that $\qthetapost$, the solution to \eq{kl_minimization}, has interior
exponential family parameter $\eta_k$.  In this case, $\qthetapost$ can be
completely characterized by its mean parameters, $m := \expectq{\theta}$
\citep{wainwright2008graphical}. One can perturb the objective in
\eq{kl_minimization} in the direction of a function $f$ of the mean parameter
$m$ by some amount $t$, where $t$ is a vector with length equal to the output of
$f$:
\begin{eqnarray}\label{eq:perturbed_elbo}
q_t & := &
  \textrm{argmin}_{q \in \mathcal{Q}} \left\{S - L + f(m)^T t \right\}
\end{eqnarray}
\citep{giordano:2015:lrvb} showed that when $f(m) = m$, we can calculate the
local change in the mean of $q_t$ as $t$ varies:
\begin{eqnarray}\label{eq:basic_lrvb}
\left. \frac{d\mathbb{E}_{q_t}\left[\theta\right]}{dt^T} \right|_{t=0} =
  \left(I - VH\right)^{-1} V =: \hat{\Sigma},
\quad \textrm{where } V := \cov_{\qthetapost}\left(\theta\right)
\textrm{ and } H := \frac{\partial^2 L}{\partial m \partial m^T}.
\end{eqnarray}
As shown in \app{functions}, if $f(m)$ and $h(m)$ are both smooth functions of
$m$, then
\begin{eqnarray}\label{eq:function_covariance}
\frac{d h(m_t)}{dt} = \nabla h ^T \hat\Sigma \nabla f
\end{eqnarray}
\eq{basic_lrvb} is the special case of \eq{function_covariance} where $h(m) =
f(m) = m$. In \citep{giordano:2015:lrvb}, the goal was to calculate a posterior
covariance estimate $\hat{\Sigma}$.  Here, our goal is to calculate a measure of
robustness to changes in $\alpha$. Let $\alpha_t = \alpha + \Delta \alpha t$ be
the value of $\alpha$ perturbed in direction $\Delta \alpha$ by an infinitesimal
scalar amount $t$. $\Delta \alpha$ may be vector- or function-valued.  Note that
$\mbeq\left[\log(p(\theta \vert \alpha))\right]$ from \eq{kl_minimization} is a
function of $m$, since it is an expectation with respect to $q$, which is
completely parameterized by $m$.  Assuming that $p(\theta \vert \alpha)$ is a
smooth function of $\alpha$, a Taylor expansion in $\Delta \alpha t$ gives
\begin{eqnarray}\label{eq:f_definition}
\mbeq\left[\log(p(\theta \vert \alpha_t))\right] &=&
 \mbeq\left[\log(p(\theta \vert \alpha))\right] +
   \frac{d}{d\alpha^T} \mbeq\left[\log(p(\theta\vert\alpha))\right]
    \Delta \alpha t + O(t^2) \Rightarrow \nonumber\\
f(m) &:=& \frac{d}{d\alpha^T} \mbeq\left[\log(p(\theta\vert\alpha))\right]
        \Delta \alpha
\quad \textrm{ and } \quad h(m) := \expectq{g(\theta)}
\end{eqnarray}
With $f(m)$ and $h(m)$ defined as in \eq{f_definition}, \eq{function_covariance}
gives us the robustness measure \eq{local_robustness}. As in LRVB, these
derivatives are in fact the exact robustness of the variational posterior
expectations to prior perturbation.  The extent to which it represents the true
prior sensitivity depends on the extent to which the MFVB means are good
estimates of the true posterior means.

\section{Robustness measures from LRVB}\label{sec:lrvb_robustness}

We now turn to calculating $f(m)$ from \eq{f_definition} for some common
cases. For simplicity, we will take $g(\theta) = \theta$.
First, consider a prior in the exponential family with sufficient statistics
$\pi(\theta)$.
\begin{eqnarray}\label{eq:finite_dim_perturbation}
\log p(\theta \vert \alpha) &=& \alpha^T \pi(\theta) \Rightarrow
f(m) = \expectq{\pi(\theta)} \Delta \alpha
\end{eqnarray}
Here, $\pi(\theta)$ is a vector of the same length as $\alpha$.
Note that $f(m)$ may be known exactly or estimated using Monte Carlo
simulation.  The simplest case is when the priors are conditionally conjugate
for $p(x \vert \theta)$.  In that case, $\pi(\theta) = \theta$, and
$\frac{d \expectq{\theta_i}}{d \alpha_j} = \hat\Sigma_{ij}$.
A more complex non-conjugate example is the LKJ prior on a covariance
matrix, which we explore in \mysec{experiments}.

Next, we consider changing the functional form of $p(\theta \vert \alpha)$,
taking $\Delta \alpha$ to be function-valued.  We will focus on perturbations to
the prior marginals, since local robustness properties of functional
neighborhoods of the full posterior have bad asymptotic properties
\citep{gustafson:1996:localposterior}. Let $\theta_i$ be a subvector of $\theta$
whose marginal we will perturb. We assume that both the prior and variational
distribution factor across $\theta_i$:
\begin{eqnarray*}
\qthetapost(\theta) = q(\theta_i) q(\theta_{-i})
\quad\textrm{ and }\quad
p(\theta \vert \alpha) =
  p(\theta_i \vert \alpha_i) p(\theta_{-i} \vert \alpha_{-i})
\end{eqnarray*}
where $-i$ denotes ${1,...,K} \setminus i$.
For simplicity of notation, assume without loss of generality
that the $i$ indices come first: $\theta^T = (\theta_i^T, \theta_{-i}^T)$
(Both $\qthetapost$ and the prior may factorize still further.)

In order to ensure that the perturbed prior is properly normalized,
we will shift an infinitesimal amount of prior mass from the original
$p(\theta_i \vert \alpha)$ to a density $p_c(\theta_i)$:
\begin{eqnarray}\label{eq:epsilon_contamination}
p(\theta_i \vert \alpha_i, \epsilon) =
  (1 - \epsilon) p(\theta_i \vert \alpha_i) + \epsilon p_c(\theta_i)
\end{eqnarray}
This is known as $\epsilon$-contamination, and
its construction guarantees that the perturbed prior is properly normalized
\footnote{$\epsilon$-contamination is principally adopted for analytic
convenience, though it is an expressive class of perturbations
\citep{gustafson:1996:localposterior}.  For more exotic perturbation classes,
which we do not consider here, see\citep{zhu:2011:bayesian}.}. By taking
$p_c(\theta_i) = \delta(\theta_i - \theta_{i0})$ to be a Dirac
delta function at $\theta_{i0}$, \eq{function_covariance} and
\eq{f_definition} give (see \app{robust_derivations}):
\begin{eqnarray}\label{eq:delta_function_sensitivity}
\frac{d \mbeq[\theta]}{ d \epsilon} &=&
  \frac{\qthetapost(\theta_{i0})}{p(\theta_{i0} \vert \alpha)}
  (I - VH)^{-1}
  \left(\begin{array}{c}
   \theta_{i0} - m_{i}\\
   0
  \end{array}\right)
\end{eqnarray}
This is known as an ``influence function''
\citep{gustafson:2012:localrobustnessbook}.
Note that $p(\theta_{i0} \vert \alpha)$ is known \textit{a priori}, and that
$\qthetapost(\theta_{i0})$ is a function of the moment parameters $m$, since $m$
entirely specifies $\qthetapost$.  Viewed as a function of $\theta_0$,
\eq{delta_function_sensitivity} characterizes how much each moment parameter,
$m$, is affected by adding an infinitesimal amount of prior mass at
$\theta_{i0}$.  By the linearity of the
derivative, one can use weighted combinations of delta functions and
\eq{delta_function_sensitivity} to estimate the sensitivity to any prior
function
\footnote{A closed form for $p_c(\theta_i)$ other than weighted
combinations of Dirac delta functions is given in \app{function_sensitivity}.
The influence function is closely related to the worst-case prior perturbation
within a metric ball in the space of prior functions
\citep{gustafson:1996:localposterior}.  We show in \app{extreme} that LRVB also
gives a closed form for this worst-case perturbation.
\app{mcmc_intuition_comparison} provides some intuition by comparing the LRVB
results to the corresponding formulas for exact inference.}.


\section{Experiments}\label{sec:experiments}



We applied the methods above to a hierarchical model of microcredit
interventions in development economics \citep{meager:2015:microcredit}. One
output of the model is $\mu$ and $\tau$, top level parameters in a hierarchical
model that measure average site profitability and the effectiveness of
microcredit interventions, respectively.  Here, we present the sensitivity of
these parameters to $\Lambda$, the information matrix of a normal prior on
$(\mu,\tau)$, and $\eta$, the concentration parameter in a non-conjugate LKJ
prior\citep{lewandowski:2009:lkj} on the covariance of $(\mu,\tau)$.
The left panel of \fig{MicrocreditMainText} shows the estimates from
\eq{finite_dim_perturbation} normalized by the posterior standard deviation. The
results are robust to $\eta$ but extremely non-robust to $\Lambda$.  The second
panel compares the prediction of \eq{finite_dim_perturbation} to the actual
change in MCMC means to a small change in $\Lambda_{11}$.  The results match
closely. The third panel shows \eq{delta_function_sensitivity}, the influence
function of the prior for $(\mu, \tau)$  on $\tau$.  The ``X'' is the posterior
mean. Adding prior mass on only one side of the mean would be highly
influential, though it is hard to imagine such a prior representing an
\emph{a priori} belief.

We formed the LRVB estimates using JuMP\citep{JuMP:LubinDunningIJOC} and used
STAN\citep{stan-manual:2015} to generate MCMC samples.  The VB and MCMC results are
nearly identical, indicating that the assumptions necessary for LRVB
hold.  Generating one set of MCMC draws took 15 minutes, and the LRVB estimates,
including calculating all the reported sensitivity measures, took 45 seconds.
For more details, see \app{microcredit_experiment}.

\begin{knitrout}
\definecolor{shadecolor}{rgb}{0.969, 0.969, 0.969}\color{fgcolor}\begin{figure}[ht!]

{\centering \includegraphics[width=0.31\linewidth,height=0.95in]{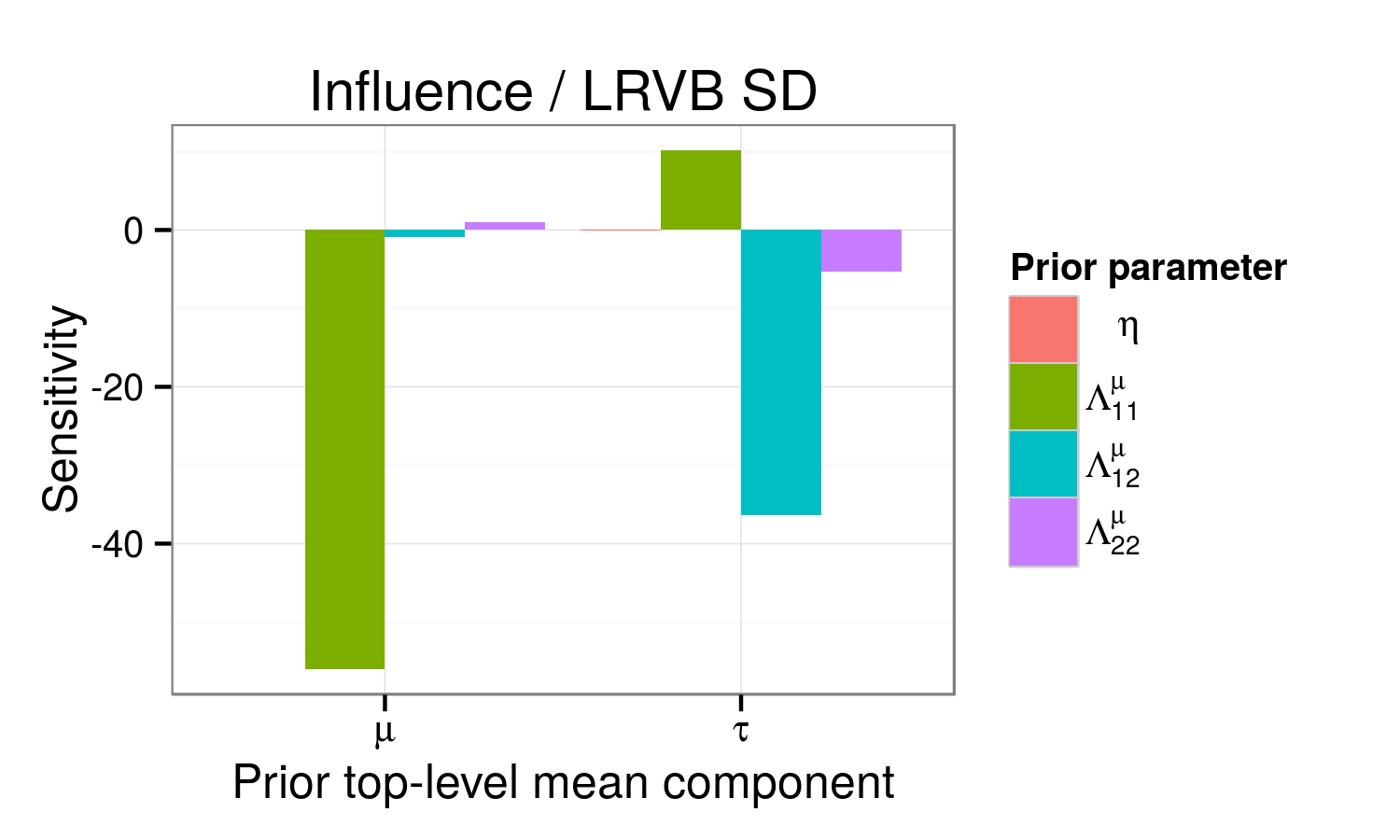}
\includegraphics[width=0.31\linewidth,height=0.95in]{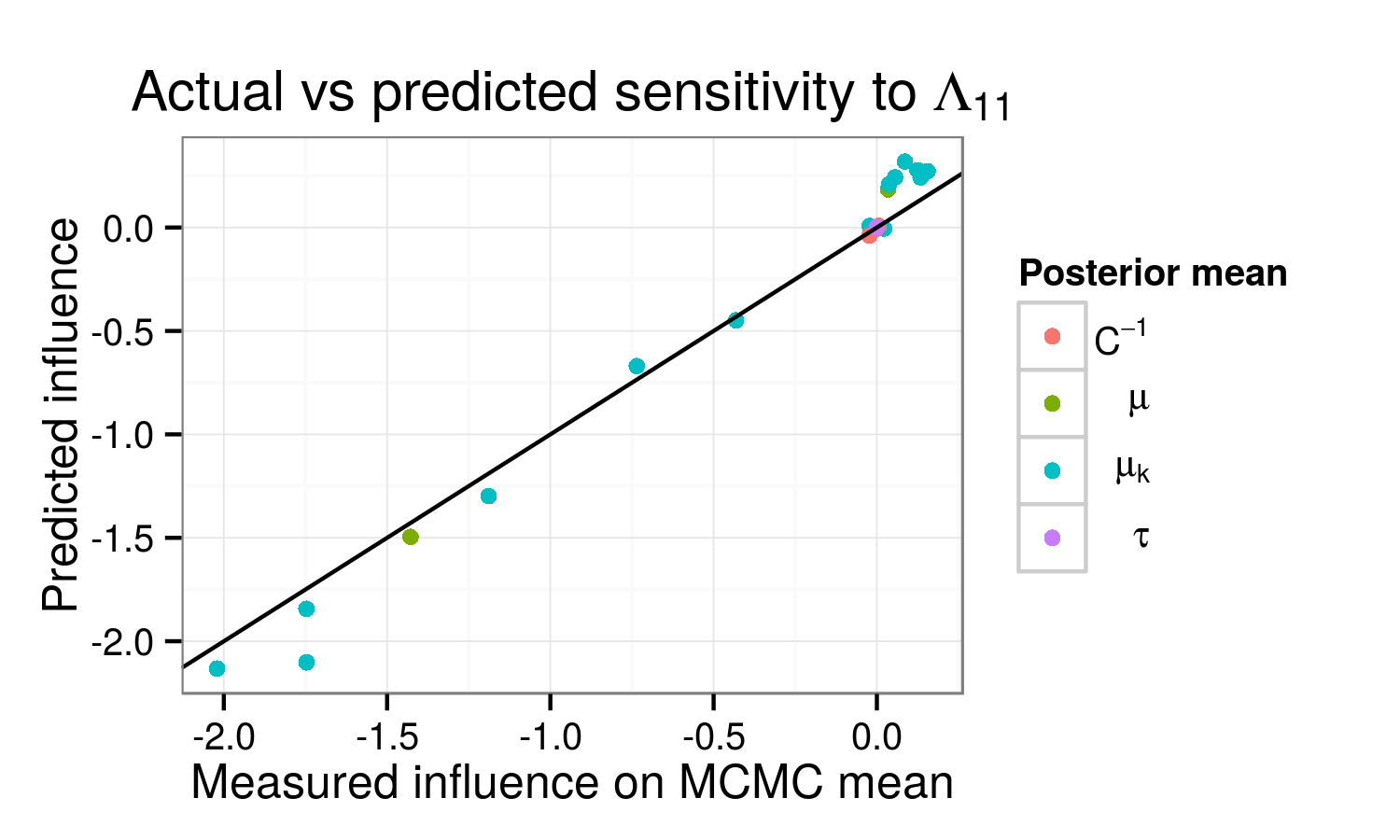}
\includegraphics[width=0.31\linewidth,height=0.95in]{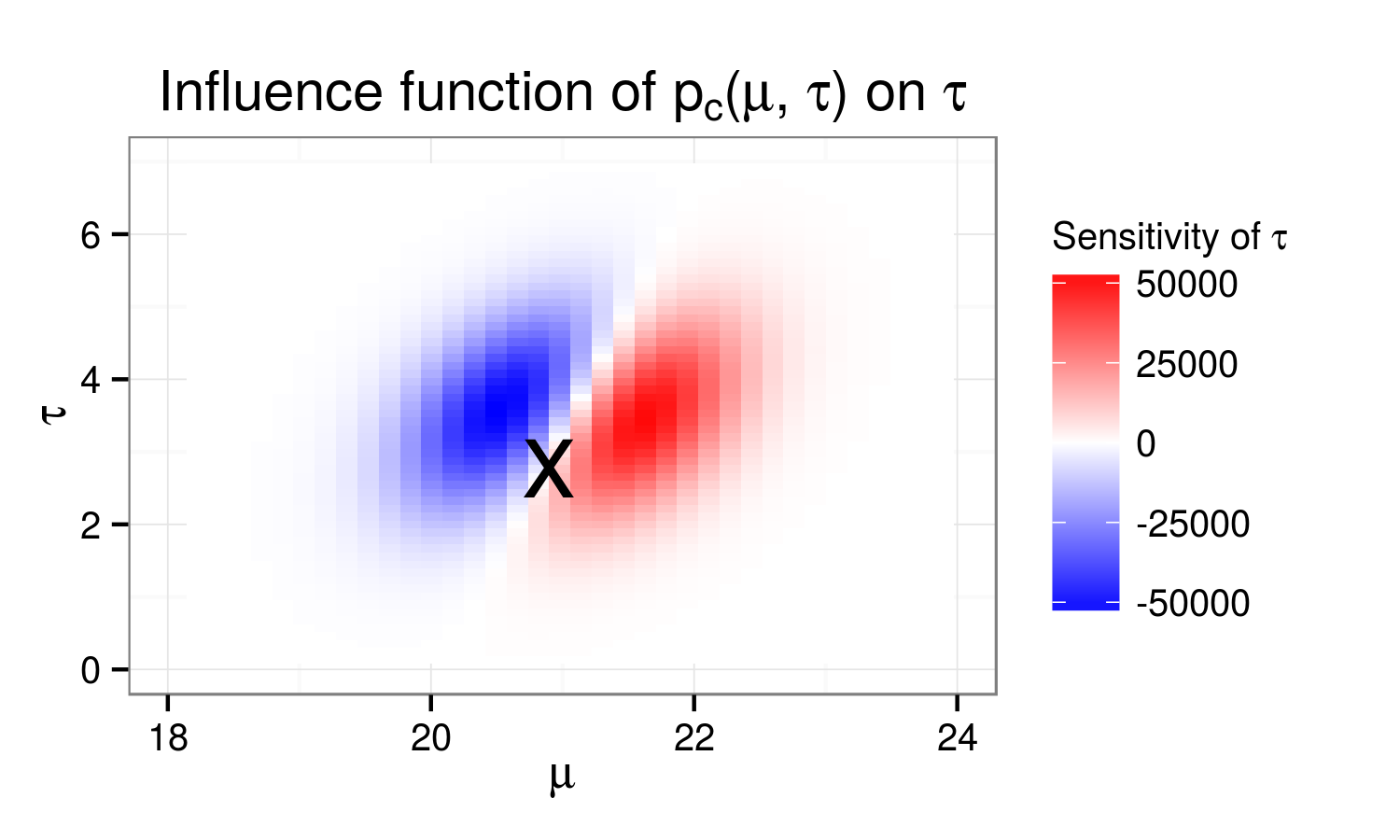}

}

\caption[Effectiveness of robustness measures in the microcredit model]{Effectiveness of robustness measures in the microcredit model}\label{fig:MicrocreditMainText}
\end{figure}

\end{knitrout}


{\small
\bibliographystyle{unsrt} 
\bibliography{influence_scores_arxiv}
}

\clearpage
\appendix
\appendixpage

\section{Robust Bayes with MCMC}\label{app:mcmc}

There is an extensive literature on Robust Bayesian techniques, surveyed
in \citep{insua:2012:robust}.  We focus on local robustness
techniques
\citep{gustafson:2012:localrobustnessbook, gustafson:1996:localposterior,
      gustafson:1996:localmarginals}.  In the original papers, many
authors focused on either theoretical results or models with special
structure that rendered robustness measures tractable.
Of MCMC, one of the founders of the field of Bayesian Robustness writes:

``The MCMC methodology was not directly compatible with many of the
robust Bayesian techniques that had been developed, so that it was
unclear how formal robust Bayesian analysis could be incorporated into the
future `Bayesian via MCMC' world.
Paradoxically, MCMC has dramatically increased the need for consideration
of Bayesian robustness, in that the modeling that is now routinely
utilized in Bayesian analysis is of such complexity that inputs (such as priors)
can be elicited only in a very casual fashion.''\citep{berger:2012:robust}

Another recent author adds:

``Surprisingly, despite considerable theoretical advances in formal
sensitivity analysis, it is barely used in every-day practice...
a formal robustness methodology which is feasible, fairly quick,
operating with low extra computing effort and provided by
default in a dedicated software, is strongly required.''
\citep{roos:2015:sensitivity}

A number of papers have proposed methods for performing robustness analyses
using MCMC techniques. \citep{gustafson:1996:localmarginals}, following many
previous theoretical works \citep{gustafson:2012:localrobustnessbook}, exchanges
the integral in a posterior expectation with the derivative with respect to
prior perturbations, giving a robustness estimate that can be evaluated from
MCMC samples. \citep{perez:2006:mcmc, perez:2006:sensitivity} extends this idea.
These approaches exploit importance sampling and / or closed forms for
derivatives or posterior densities, and care must be taken to control the
variance of the MCMC estimates. The papers
\citep{kass:1989:approximate,mcculloch:1989:local} make second-order
approximations to the log posterior and employ numerical techniques to calculate
robustness measures. \citep{roos:2015:sensitivity} uses a sophisticated
methodology to choose a grid of prior points at which they numerically estimate
the sensitivity using estimates of the posterior density.
\citep{bornn:2010:sequential} proposes a distinctive method based on particle
filtering in which particle weights are re-adjusted to produce draws from a
perturbed prior. The authors are unaware of any previous work applying robust
Bayes techniques in the context of variational methods.

The advantage of using robust Bayes with LRVB over these MCMC-based techniques
is simplicity and computational ease.  Little extra code and no extra
approximations or assumptions beyond that required for LRVB are required to
compute the robustness measures below. LRVB robustness measures are the exact
sensitivity of the variational solution to changes in the prior, and they will
be accurate to the extent that the variational approximation to the posterior
mean of interest is accurate \citep{giordano:2015:lrvb}.

\section{LRVB covariance of functions}\label{app:functions}

Let us consider LRVB estimates of the covariances of functions of
natural parameters rather than the natural parameters themselves.
Suppose we have a function $\phi\left(\eta\right)$, and a variational
solution $q\left(m\right)$ where $m=\mbeq\left[\eta\right]$. Since
$q$ is fully parameterized by $m$, we can write

\begin{eqnarray*}
\mbeq\left[\phi\left(\eta\right)\right] & = & f\left(m\right)
\end{eqnarray*}

for some continuous $f\left(m\right)$. We can consider a perturbed
log likelihood that also includes $f\left(m\right)$:
\begin{eqnarray*}
\log p_{t} & = & \log p+t_{0}^{T}m+t_{f}f\left(m\right):=\log p+t^{T}m_{f}\\
t & := & \left(\begin{array}{c}
t_{0}\\
t_{f}
\end{array}\right)\\
m_{f} & := & \left(\begin{array}{c}
m\\
f\left(m\right)
\end{array}\right)
\end{eqnarray*}

As in \citep{giordano:2015:lrvb}, we use the fixed point equations:
\begin{eqnarray*}
E_{t} & := & E+t^{T}m_{f}\\
\frac{dE_{t}}{dm} & = & 0\Rightarrow\\
\frac{dE}{dm}+\left(\begin{array}{cc}
I & \nabla f\end{array}\right)\left(\begin{array}{c}
t_{0}\\
t_{f}
\end{array}\right) & = & 0\\
M\left(m\right) & := & \frac{\partial E}{\partial m}+m\\
M_{t}\left(m\right) & := & M\left(m\right)+\left(\begin{array}{cc}
I & \nabla f\end{array}\right)\left(\begin{array}{c}
t_{0}\\
t_{f}
\end{array}\right)\\
M_{t}\left(m^{*}\right) & := & m^{*}\textrm{ (definition of }m^{*}\textrm{)}\\
\frac{dm_{t}^{*}}{dt^{T}} & = &
  \left.\frac{\partial M_{t}}{\partial m^{T}}\right|_{_{m=m_{t}^{*}}}
  \frac{dm_{t}^{*}}{dt^{T}}+\frac{\partial M_{t}}{\partial t^{T}}\\
 & = & \left(\left.\frac{\partial M}{\partial m^{T}}\right|_{_{m=m_{t}^{*}}}+
 \frac{\partial}{\partial m^{T}}\left(\begin{array}{cc}
I & \nabla f\end{array}\right)\left(\begin{array}{c}
t_{0}\\
t_{f}
\end{array}\right)\right)\frac{dm^{*}}{dt^{T}}+\left(\begin{array}{cc}
I & \nabla f\end{array}\right)
\end{eqnarray*}

The term $\frac{\partial}{\partial m^{T}}\left(\begin{array}{cc}
I & \nabla f\end{array}\right)\left(\begin{array}{c}
t_{0}\\
t_{f}
\end{array}\right)$ is awkward, but it disappears when we evaluate at $t=0$,
giving
\begin{eqnarray*}
\frac{dm_{t}^{*}}{dt^{T}} & = &
  \left(\left.\frac{\partial M}{\partial m^{T}}\right|_{_{m=m_{t}^{*}}}\right)
  \frac{dm^{*}}{dt^{T}}+\left(\begin{array}{cc}
I & \nabla f\end{array}\right)\\
 & = & \left(\frac{\partial^{2}E}{\partial m\partial m^{T}}+
  I\right)\frac{dm^{*}}{dt^{T}}+\left(\begin{array}{cc}
I & \nabla f\end{array}\right)\\
\frac{dm^{*}}{dt^{T}} & = &
  -\left(\frac{\partial^{2}E}{\partial m\partial m^{T}}\right)^{-1}
  \left(\begin{array}{cc}I & \nabla f\end{array}\right)
\end{eqnarray*}

Recalling that
\begin{eqnarray*}
\frac{dm^{*}}{dt_{0}^{T}} & := & \hat{\Sigma}
\end{eqnarray*}

We can plug in to see that

\begin{eqnarray*}
\frac{dm^{*}}{dt_{f}^{T}} & = & \hat{\Sigma}\nabla f
\end{eqnarray*}

This means that the covariance of the natural sufficient statistics
with the function $\phi\left(\eta\right)$ are determined by a linear
combination of the LRVB covariance matrix.

A similar conclusion can be reached by considering the response of
the expectation of a quantity other than a natural parameter to a
generic perturbation. Consider perturbing the log likelihood by some
function $t_{g}g\left(m\right)$. Then by the reasoning above,
\begin{eqnarray*}
\frac{df\left(m\right)}{dt_{g}} & = & \frac{df}{dm^{T}}\frac{dm}{dt_{g}}
  =  \nabla f^{T}\hat{\Sigma}\nabla g
\end{eqnarray*}

This is \eq{function_covariance}, and represents the LRVB covariance between two
quantities with variational expectation $f\left(m\right)$ and $g\left(m\right)$
respectively.  As in the present, that covariance can also be interpreted as the
sensitivity of $g(m)$ to a perturbation of the objective by $g(m)$.

\section{Robustness Derivations} \label{app:robust_derivations}

In this section, we derive results stated in \mysec{lrvb_robustness}.
For generality, when possible we will derive results for the full
vector $\theta$ rather than the sub-vector $\theta_i$ when the proof
would be identical for the subvector under the assumption that
$q(\theta) = q(\theta_i) q(\theta_{-i})$.

\subsection{Sensitivity to $\epsilon-$contamination}

For a given $p_c(\theta)$ in \eq{epsilon_contamination}, we can consider
$p(\theta \vert \alpha, \epsilon)$ to be a class of priors parameterized
by $(\alpha, \epsilon)$, and take $\epsilon = \Delta \alpha$ in
\eq{f_definition}.  We then need to calculate

\begin{eqnarray*}
\left. \frac{d}{d\epsilon}
  \mbeq\left[\log p(\theta \vert \alpha, \epsilon) \right] \right|_{\epsilon = 0} &=&
\mbeq\left[\left. \frac{d}{d\epsilon}
  \log \left((1 - \epsilon) p(\theta \vert \alpha) + \epsilon p_c(\theta)\right)
    \right|_{\epsilon = 0} \right] \\
&=& \mbeq\left[\frac{p_c(\theta)}{p(\theta \vert \alpha)} - 1\right] \\
\end{eqnarray*}

Since the variational solution is unaffected by adding constants to the ELBO,
we can take

\begin{eqnarray}\label{eq:epsilon_f_formula}
f(m) &:=& \mbeq\left[\frac{p_c(\theta)}{p(\theta \vert \alpha)}\right]
\end{eqnarray}

\subsection{Sensitivity to a function}\label{app:function_sensitivity}

We will calculate $\nabla f(m)$ using \eq{epsilon_f_formula} for a general
function $p_c(\theta)$ and then use this result to derive
\eq{delta_function_sensitivity} as a special case.  In this section,
we rely on the fact that the variational distribution, $q(\theta)$, is
in the exponential family.

The directional derivative for a perturbation $p_c(\theta)$ is given by
the Taylor expansion of $\mbeq\left[\frac{p_c(\theta)}{p\left(\theta
\vert \alpha\right)}\right]$ in terms of the exponential family moment
parameters:
\begin{eqnarray}\label{eq:exepcted_general_function_perturb}
\frac{d}{dm}\mbeq\left[\frac{p_c(\theta)}{p\left(\theta \vert \alpha\right)}\right]	&=&
  V^{-1}\frac{d}{d\eta}\int\exp\left(\eta^{T}\theta-A\left(\theta\right)\right)
    \frac{p_c(\theta)}{p\left(\theta \vert \alpha\right)}d\theta \nonumber\\
 &=&	V^{-1}\int q\left(\theta\right)\left(\theta-m\right)
      \frac{p_c(\theta)}{p\left(\theta \vert \alpha\right)}d\theta \nonumber\\
 &=&	V^{-1}\mbeq\left[\left(\theta-m\right)
      \frac{p_c(\theta)}{p\left(\theta \vert \alpha\right)}\right]
\end{eqnarray}

Taking $p_c(\theta) = \delta\left(\theta_{i} = \theta_{i0}\right)$ to
be a Dirac delta function gives \eq{delta_function_sensitivity}.

\subsection{Extremal derivative}\label{app:extreme}

The influence function is closely related to the worst-case prior perturbation
in a metric ball around the original prior, $p(\theta_i \vert \alpha_i)$. We
refer the reader to \citep{gustafson:1996:localposterior} for the background.
Given \eq{exepcted_general_function_perturb}, the proof for the variational case
is essentially identical.

First, to match \citep{gustafson:1996:localposterior}, let $p_c(\theta)$ be a
signed measure and consider perturbations of the form
\begin{eqnarray*}
p(\theta \vert \alpha, \epsilon) = p(\theta \vert \alpha) + \epsilon p_c(\theta)
\end{eqnarray*}
Because the variational solution is invariant to constants, the variational
sensitivity to this perturbation is identical to that of $\epsilon-$
contamination.  Consequently, the sensitivity is given by
\eq{exepcted_general_function_perturb} and \eq{function_covariance}:
\begin{eqnarray*}
\frac{d\mbeq\left[g(\theta)\right]}{dt} & = &
  \nabla h ^{T}\hat{\Sigma}V^{-1}\mbeq\left[\left(\theta-m\right)
  \frac{p_c(\theta)}{p\left(\theta \vert \alpha\right)}\right]\\
 & = & \mbeq\left[\nabla h ^{T} \left(I-VH\right)^{-1}\left(\theta-m\right)
  \frac{p_c(\theta)}{p\left(\theta \vert \alpha\right)}\right]
\end{eqnarray*}

Define
\begin{eqnarray*}
a\left(\theta\right) & = &
  \nabla h ^{T} \left(I-VH\right)^{-1}\left(\theta-m\right)
  \frac{q\left(\theta\right)}{p\left(\theta \vert \alpha\right)}
\end{eqnarray*}

As in \citep{gustafson:1996:localposterior},
for $p \in [1, \infty]$ and $\frac{1}{p}+\frac{1}{q}=1$, define the size
of a perturbation as
\begin{eqnarray} \label{eq:perturb_size}
\left(\int \left| \frac{ p_c(\theta)}{p(\theta \vert \alpha)} \right| ^p
  d\Pi\right)^{\frac{1}{p}}
\end{eqnarray}
...where $\Pi$ is the measure on $\theta$ induced by $p(\theta \vert \alpha)$
and $p \in [1, \infty]$.  Let $(\cdot)^{+}$ denote the positive part and
$(\cdot)^{-}$ the negative part of the term in the parentheses.

\begin{eqnarray*}
\mbeq\left[\left|R^{T}\left(I-VH\right)^{-1}\left(\theta-m\right)
  \frac{q\left(\theta\right)}{p\left(\theta \vert \alpha\right)}\right|^{+}\right] & = &
  \int\left|a\left(\theta\right)^{+}
  \frac{p_c(\theta)}{p\left(\theta \vert \alpha\right)}\right|d\Pi\\
 & \le & \left(\int\left|a\left(\theta\right)^{+}\right|^{q}
    d\Pi\right)^{\frac{1}{q}}
  \left(\int\left|\frac{p_c(\theta)}{p\left(\theta \vert \alpha\right)}
  \right|^{p}d\Pi\right)^{\frac{1}{p}}\\
 & = & \left(\int\left|a\left(\theta\right)^{+}\right|^{q}d\Pi\right)^{\frac{1}{q}}
\end{eqnarray*}

Since we are taking $p_c(\theta)$ such that
$\|\frac{p_c(\theta)}{p\left(\theta \vert \alpha\right)};\Pi\|_{p}=1$.
This is maximized when
\begin{eqnarray*}
\left|a\left(\theta\right)^{+}\right|^{q} & \propto &
  \left|\frac{p_c(\theta)}{p\left(\theta \vert \alpha\right)}\right|^{p}\\
p_c(\theta) & = & \pi\left|\left(R^{T}\left(I-VH\right)^{-1}
  \left(\theta-m\right)\right)^{+}
  \frac{q\left(\theta\right)}{p\left(\theta \vert \alpha\right)}\right|^{\frac{q}{p}}
\end{eqnarray*}

A similar analysis follows for $a(\theta)^{-}$, and it follows that the
worst-case prior perturbation in a $p-$neighborhood of $p(\theta \vert \alpha)$
is given by

\begin{eqnarray}\label{eq:extremal_perturbation}
p_c(\theta) &=&
  p(\theta \vert \alpha)
  \max\left\{\left|a(\theta)^{+}\right|^{\frac{1}{p-1}},
            \left|a(\theta)^{-}\right|^{\frac{1}{p-1}}  \right\}
\end{eqnarray}

\subsection{Comparison with Exact Results}\label{app:mcmc_intuition_comparison}

Comparing \eq{extremal_perturbation} with
\citep[Equation~6]{gustafson:1996:localposterior} lends some intuition.
In our notation, the exact extremal perturbation is given by
\eq{gustafson_extremal} by the same expression as \eq{extremal_perturbation}
but with a different $a(\theta)$:
\begin{eqnarray}\label{eq:gustafson_extremal}
  a_p\left(\theta\right) &=&
    g(\theta) \left(\theta - \expectp{\theta} \right)
    \frac{p\left(\theta \vert x\right)}{p(\theta \vert \alpha)}
\end{eqnarray}
Here, $\qthetapost$ plays the role of the marginal posterior $p(\theta \vert x)$,
and  $\expectq{g(\theta)}^T (I-VH)^{-1}$ plays the role of $g(\theta)$.
Note that a principal difficulty of using
\eq{gustafson_extremal} is that \eq{gustafson_extremal}
requires knowledge of ratio of the posterior density to the prior
density, which is not automatically available from MCMC draws.  The
MFVB solution circumvents this difficulty by providing an explicit parametric
approximation to the posterior density.

\section{Microcredit Model}\label{app:microcredit_experiment}

\newcommand{\mcPriorEta}{15.01}
\newcommand{\mcPriorMuInfo}{0.02}
\newcommand{\mcPriorScaleAlpha}{20.01}
\newcommand{\mcPriorScaleBeta}{20.01}
\newcommand{\mcPriorTauAlpha}{2.01}
\newcommand{\mcPriorTauBeta}{2.01}

We will reproduce a variant of the analysis performed in
\citep{meager:2015:microcredit}, though with somewhat different prior choices.
Randomized controlled trials were run in seven different sites to try to measure
the effect of access to microcredit on various measures of business success.
Each trial was found to lack power individually for various reasons, so there
could be some benefit to pooling the results in a simple hierarchical model. For
the purposes of demonstrating robust Bayes techniques with VB, we will focus on
the simpler of the two models in \citep{meager:2015:microcredit} and ignore
covariate information.

We will index sites with $k=1,..,K$ (here, $K=7$) and business within a site by
$i=1,...,N_k$.  In site $k$ and business $i$ we observe whether the business was
randomly selected for increased access to microcredit, denoted $T_{ik}$, and the
profit after intervention, $y_{ik}$.  We follow \citep{rubin:1981:estimation} and
assume that each site has an idiosyncratic average profit, $\mu_k$ and average
improvement in profit, $\tau_k$, due to the intervention. Given $\mu_k$,
$\tau_k$, and $T_{ik}$, the observed profit is assumed to be generated according
to

\begin{eqnarray*}
y_{ik} \vert \mu_k, \tau_k, T_{ik}, \sigma_{k} &\sim&
  N\left(\mu_k + T_{ik} \tau_k, \sigma^2_{k} \right)
\end{eqnarray*}

The site effects, $(\mu_k, \tau_k)$, are assumed to come from an overall
pool of effects and may be correlated:

\begin{eqnarray*}
\left( \begin{array}{c} \mu_k \\ \tau_k \end{array}\right) &\sim&
N\left(
  \left( \begin{array}{c} \mu \\ \tau \end{array}\right), C \right) \\
C &:=&
\left( \begin{array}{cc}
  \sigma^2_\mu & \sigma_{\mu\tau} \\
  \sigma_{\mu\tau} & \sigma^2_\tau
\end{array}\right)
\end{eqnarray*}

The effects $\mu$, $\tau$, and the covariance matrix $V$ are unknown parameters
that require priors. For $(\mu, \tau)$ we simply use a bivariate normal prior.
However, choosing an appropriate prior for a covariance matrix can be
conceptually difficult \citep{barnard:2000:modeling}. Following the recommended
practice of the software package STAN\citep{stan-manual:2015}, we derive a
variational model to accommodate the non-conjugate LKJ prior
\cite{lewandowski:2009:lkj}, allowing the user to model the covariance and
marginal variances separately. Specifically, we use

\begin{eqnarray*}
C &=:& SRS\\
S &=& \textrm{Diagonal matrix}\\
R &=& \textrm{Covariance matrix}\\
S_{kk} &=& \sqrt{\textrm{diag}(C)_k}\\
\end{eqnarray*}

We can then put independent priors on the scale of the variances, $S_{kk}$,
and on the covariance matrix, $R$.  We model the inverse of $C$ with a Wishart
variational distribution, and use the following priors:

\begin{eqnarray*}
q\left(C^{-1}\right) &=& \textrm{Wishart}(V_\Lambda, n)\\
p\left(S\right) &=& \prod_{k=1}^{2} p(S_{kk})\\
S_{kk}^{2} &\sim& \textrm{InverseGamma}(\alpha_{scale}, \beta_{scale})\\
\log p(R) &=&  (\eta - 1) \log |R| + \constant\\
\end{eqnarray*}

The necessary expectations have closed forms with the Wishart variational
approximation, as derived in \app{lkj}.

In addition, we put a normal prior on $(\mu, \tau)^T$ and an inverse
gamma prior on $\sigma_k^2$:

\begin{eqnarray*}
\left(\begin{array}{cc} \mu \\ \tau \end{array}\right)
    &\sim& N \left(\left(\begin{array}{cc} 0 \\ 0 \end{array}\right), \Lambda^{-1} \right)\\
\sigma_{k}^{2} &\sim&
  \textrm{InverseGamma}(\alpha_\tau, \beta_\tau) \\
\end{eqnarray*}

The prior parameters used were:
\begin{eqnarray*}
  \Lambda &=& \left( \begin{array}{cc}
  \mcPriorMuInfo & 0 \\
  0 & \mcPriorMuInfo
  \end{array}\right)  \\
  \eta &=& \mcPriorEta \\
  \sigma_{k}^{-2} &\sim&
    \textrm{InverseGamma}(\mcPriorTauAlpha, \mcPriorTauBeta) \\
  \alpha_{scale} &=& \mcPriorScaleAlpha \\
  \beta_{scale} &=& \mcPriorScaleBeta \\
  \alpha_{\tau} &=& \mcPriorTauAlpha \\
  \beta_{\tau} &=& \mcPriorTauBeta
\end{eqnarray*}

\subsection{Results}

First, note that the the MCMC results match the VB means very closely,
indicating that the assumptions underlying LRVB are satisfied.  The least-
well estimated parameters are $C^{-1}$.

\begin{knitrout}
\definecolor{shadecolor}{rgb}{0.969, 0.969, 0.969}\color{fgcolor}\begin{figure}[ht!]

{\centering \includegraphics[width=0.49\linewidth,height=0.3\linewidth]{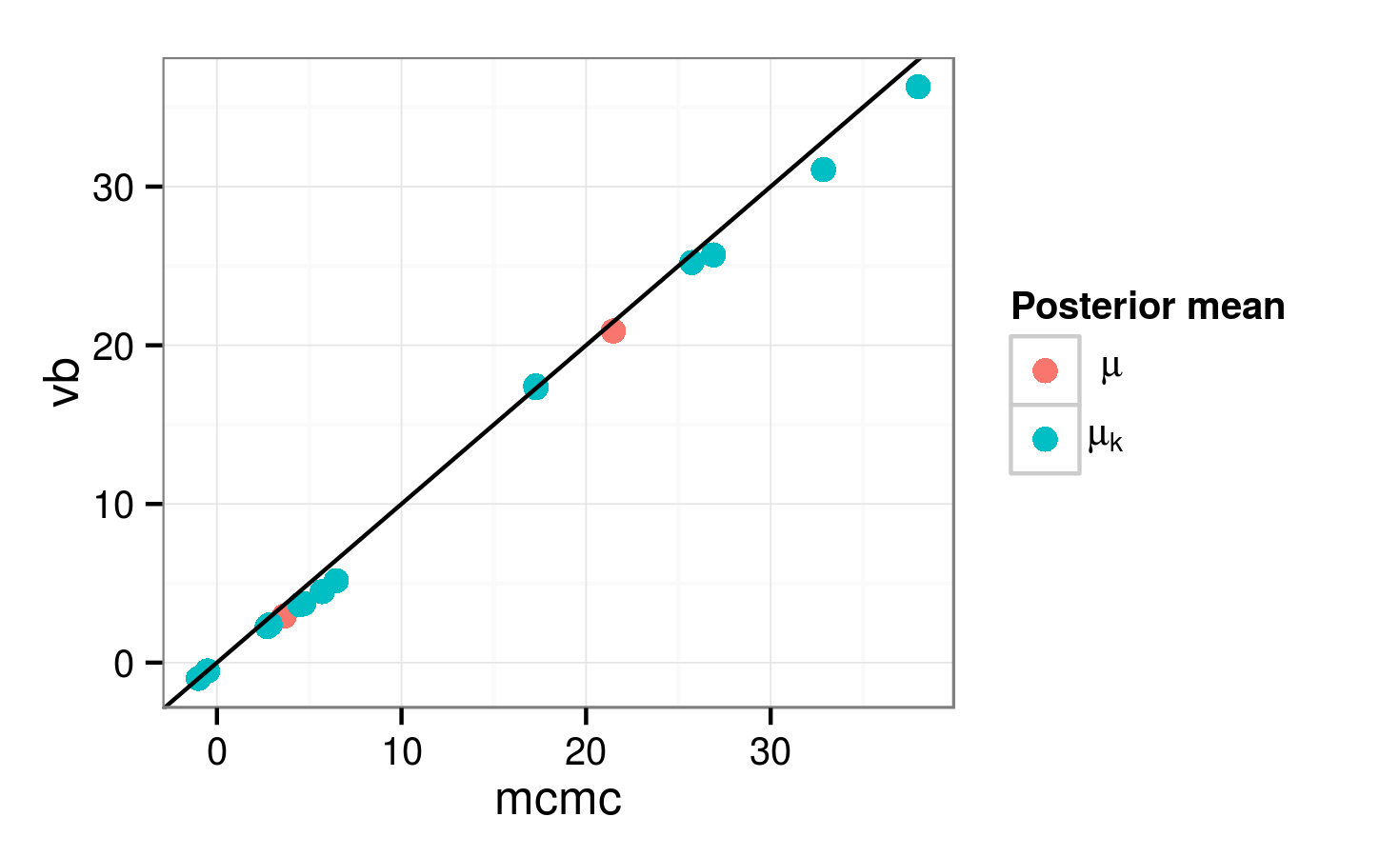}
\includegraphics[width=0.49\linewidth,height=0.3\linewidth]{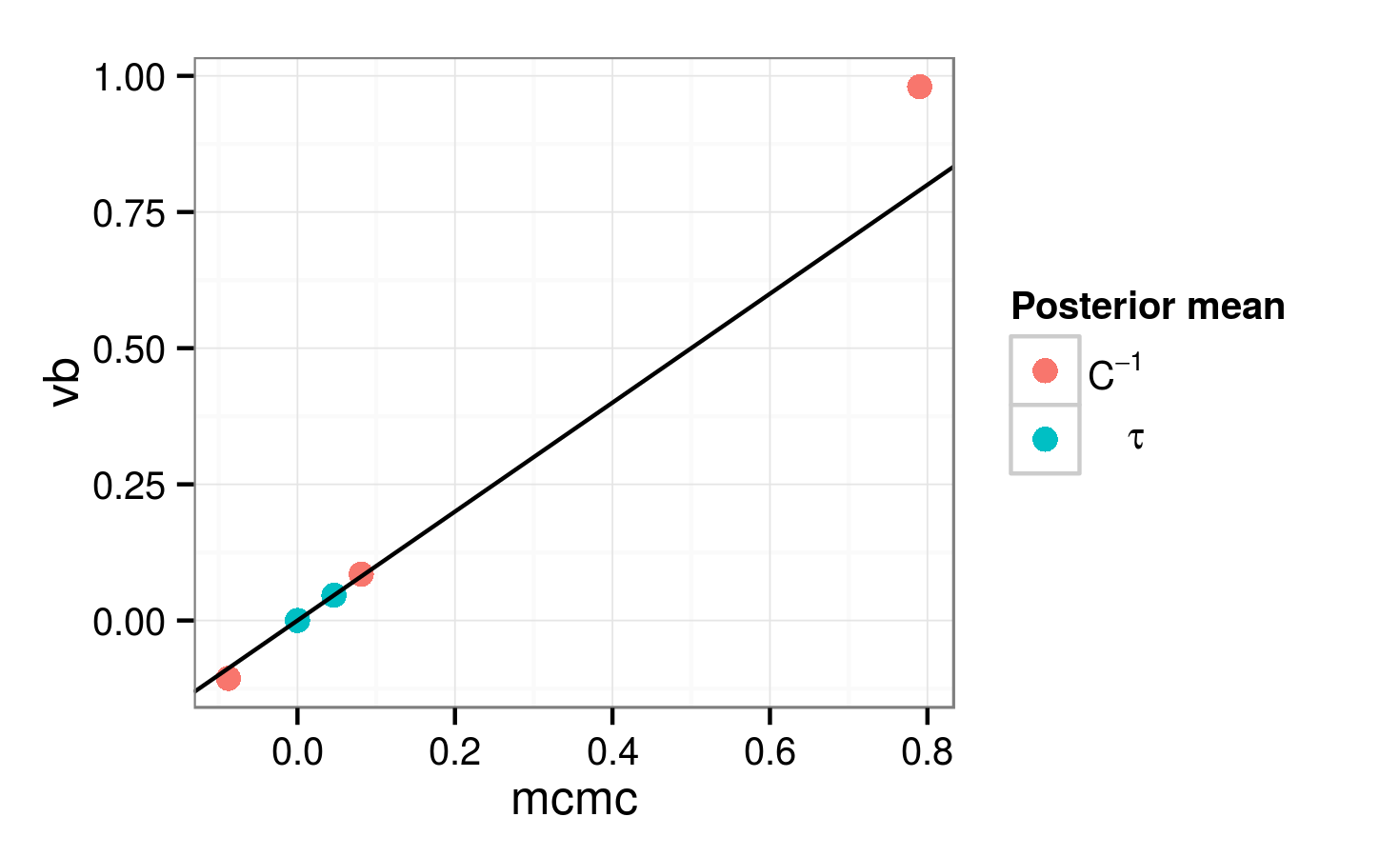}

}

\caption[Comparison of MCMC and VB Results for the microcredit data]{Comparison of MCMC and VB Results for the microcredit data}\label{fig:MicrocreditMCMCComparison}
\end{figure}

\end{knitrout}

We will focus on the robustness of $\mu$ and $\tau$, since as the higher-level
parameters in the hierarchical model, they are both more susceptible to prior
influence and more generally interpretable (as the average profit and
the causal effect of microcredit, respectively).
The sensitivity of $(\mu, \tau)$ to $\Lambda$ and $\eta$ is shown in
the left panel of \fig{MuGraph} as a proportion of the LRVB posterior
standard deviation.  The parameters can be seen to be quite sensitive
to changes in $\Lambda$.  For example, if the upper left component
of $\Lambda$, $\Lambda_{11}$, were to increase by $0.04$, $\mbeq[\mu]$ would
be expected to increase by two posterior standard deviations.  If $0.06$
is a subjectively reasonable value for $\Lambda_{11}$, then the ordinary
posterior confidence interval for $\mu$ is quite inadequate in capturing
the subjective range of beliefs that might be assigned to $\mu$.  In contrast,
the sensitivty to $\eta$, the LKJ parameter, is quite small.

The right panel of \fig{MuGraph} shows the influence function of $(\mu, \tau)$
on $\tau$.  The $X$ marks the posterior mean.  Recall that the prior mean
is $(0,0)$ and relatively diffuse.  The numbers are quite large,
indicating that adding a small amount of prior mass precisely near the
posterior could influence the posterior considerably.  However, such
a prior perturbation would have to have informed by the data -- adding
mass nearly anywhere else would have a much smaller effect.  What kind
of prior perturbation is reasonable remains a subjective decision of the
modeler.

\begin{knitrout}
\definecolor{shadecolor}{rgb}{0.969, 0.969, 0.969}\color{fgcolor}\begin{figure}[ht!]

{\centering \includegraphics[width=0.49\linewidth,height=0.3\linewidth]{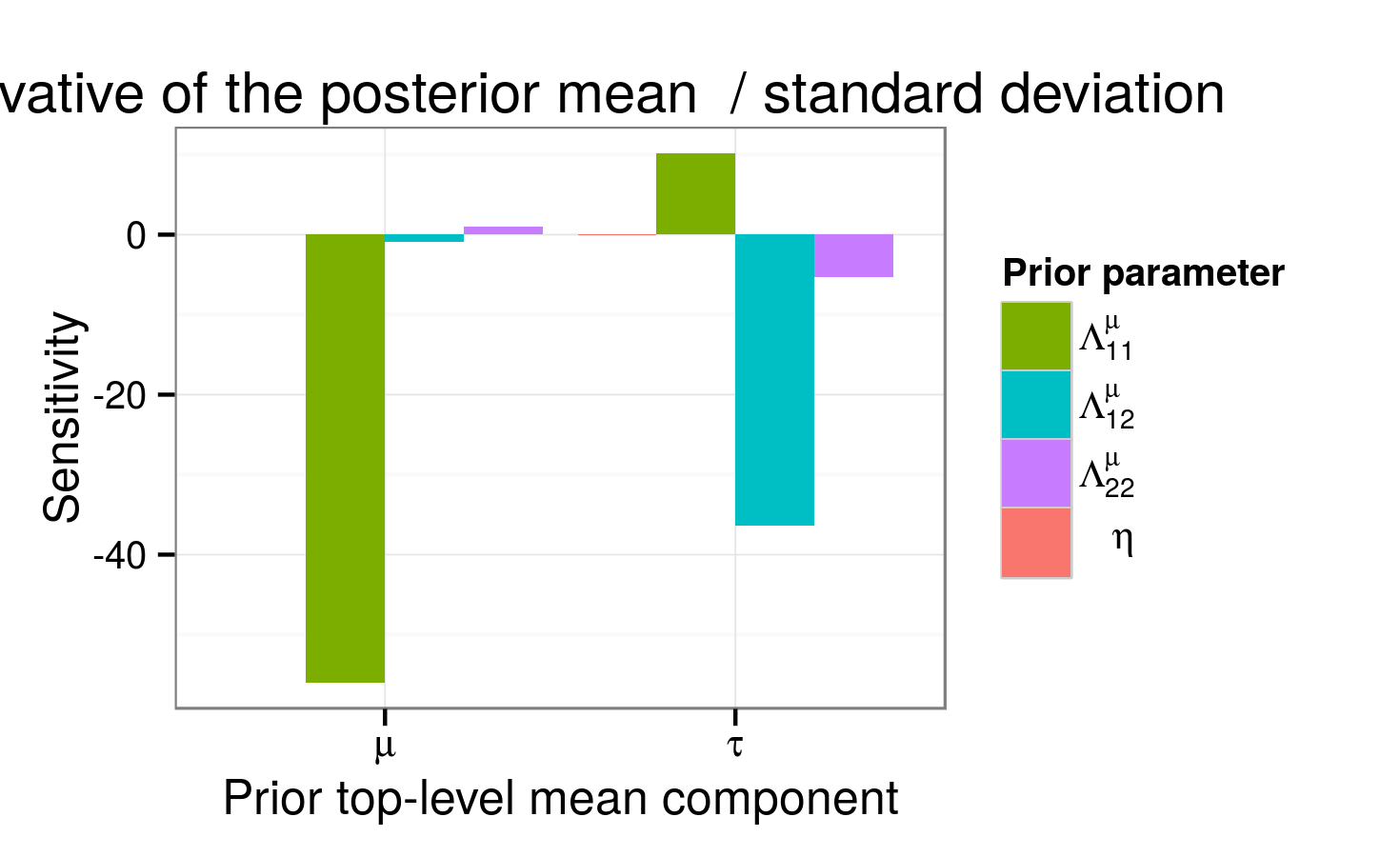}
\includegraphics[width=0.49\linewidth,height=0.3\linewidth]{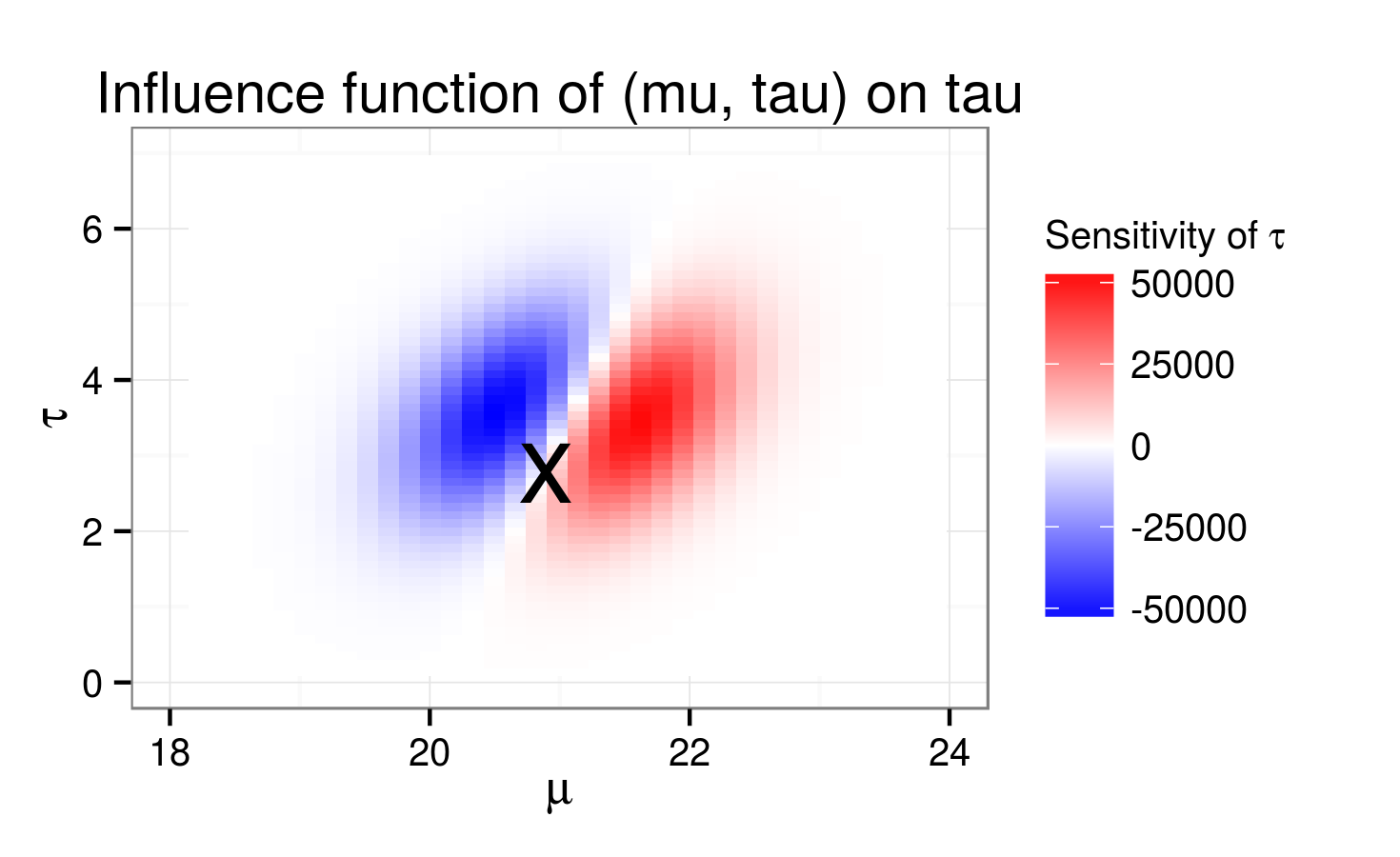}

}

\caption[The sensitivity of ]{The sensitivity of $\mu$ and $\tau$}\label{fig:MuGraph}
\end{figure}

\end{knitrout}

Finally, \fig{MicrocreditPerturbation} shows the effects of changing
$\Lambda_{11}$ on a re-run MCMC chain compared with the effects predicted
by LRVB robustness measurements.  The results are very good for all except
$C^{-1}$, which was not estimated well by the VB model.  Even for
$C^{-1}$, the LRVB estimates are directionally correct.

\begin{knitrout}
\definecolor{shadecolor}{rgb}{0.969, 0.969, 0.969}\color{fgcolor}\begin{figure}[ht!]

{\centering \includegraphics[width=0.49\linewidth,height=0.3\linewidth]{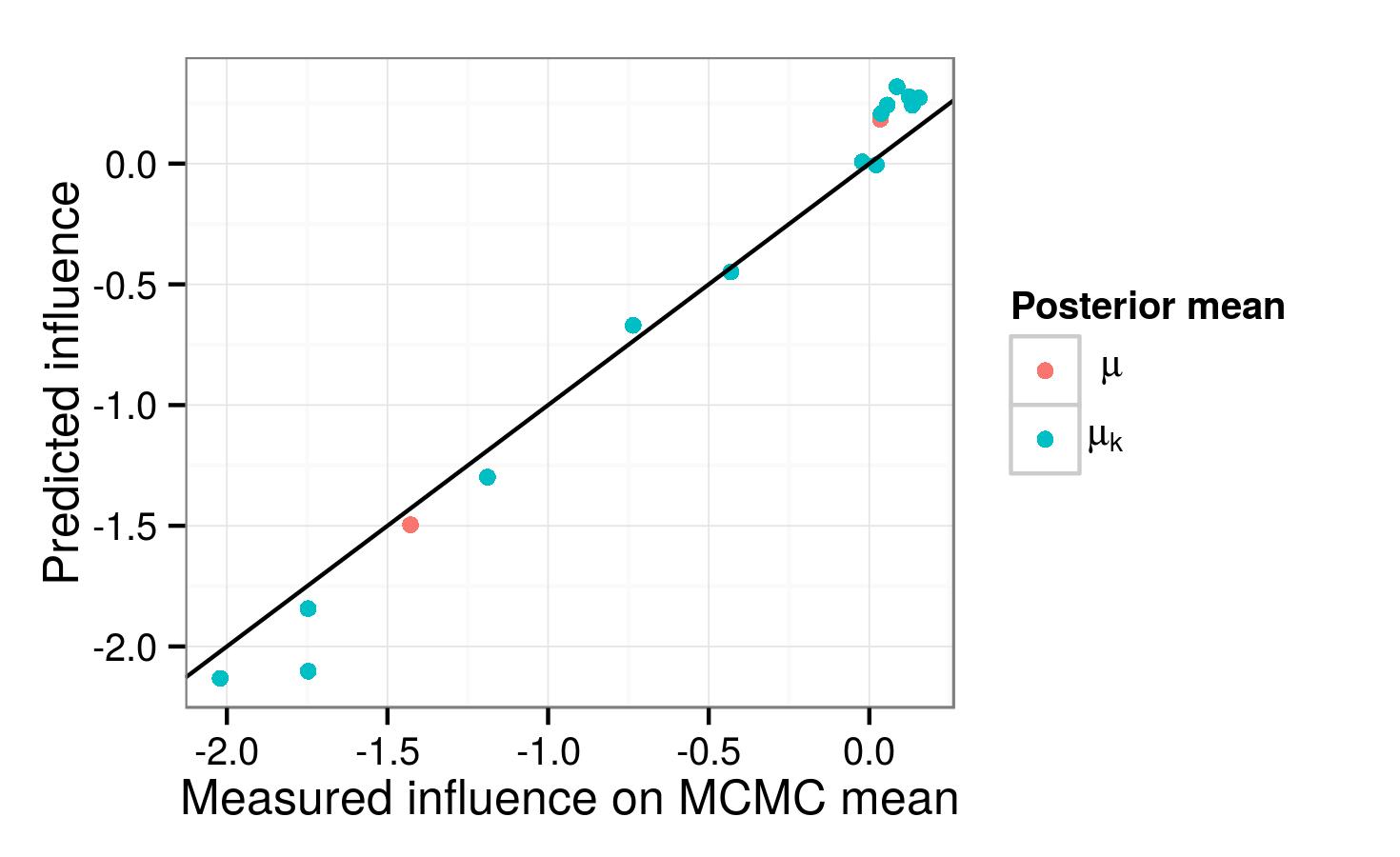}
\includegraphics[width=0.49\linewidth,height=0.3\linewidth]{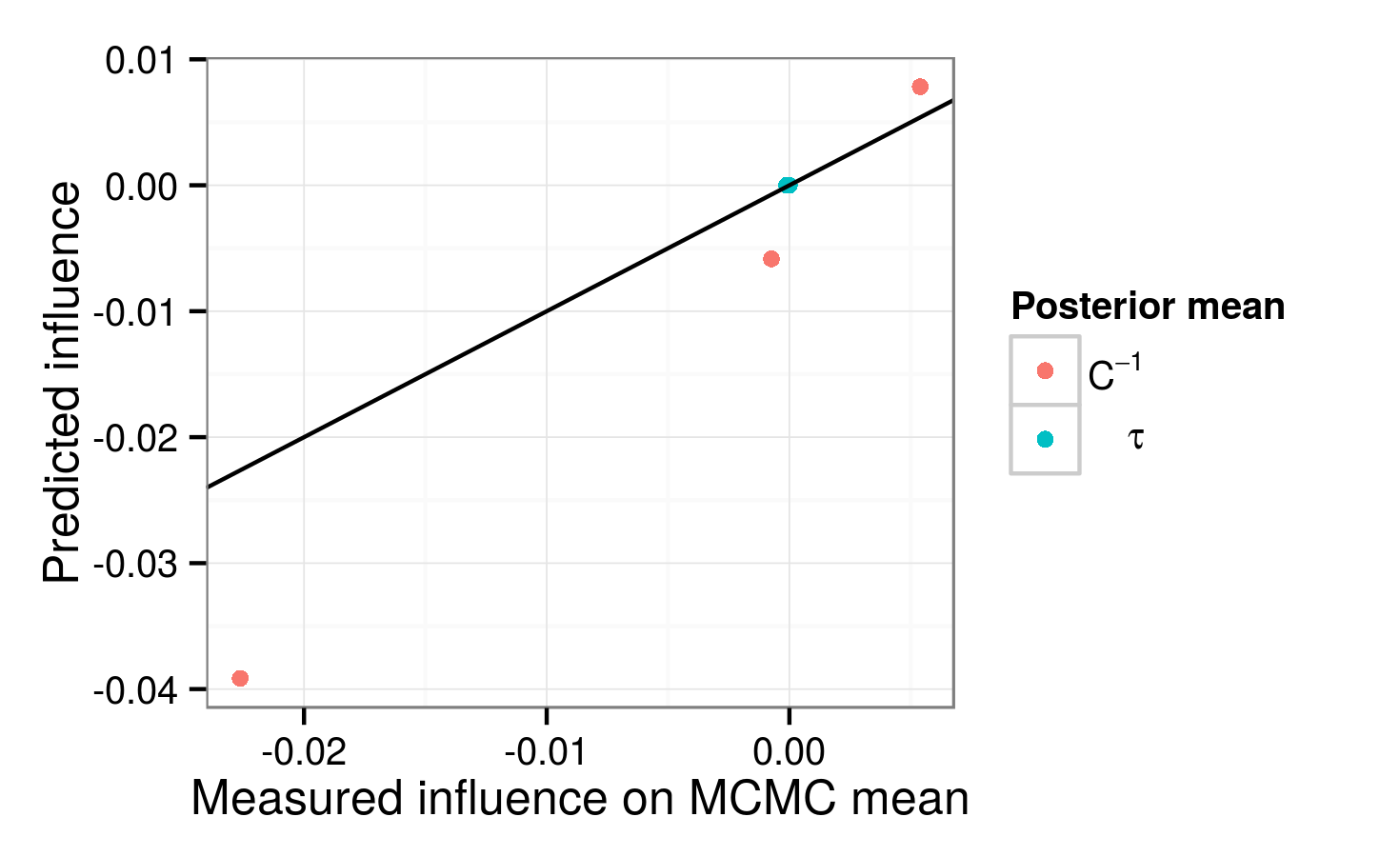}

}

\caption[Predicted vs Actual effects of perturbations]{Predicted vs Actual effects of perturbations}\label{fig:MicrocreditPerturbation}
\end{figure}

\end{knitrout}

\section{LKJ Priors for Covariance Matrices in Mean Field Variational Inference}\label{app:lkj}

In this section we briefly derive closed form expressions for using an
LKJ prior with a Wishart variational approximation.

We want to estimate a multivariate normal covariance matrix with flexible
priors. For simplicity, let us study in isolation the model:
\begin{eqnarray*}
\log p\left(y\vert\Lambda\right) & = & -\frac{1}{2}y^{T}\Lambda y+\frac{1}{2}\log\Lambda\\
\Lambda & = & \Sigma^{-1}\\
\Sigma & =: & SRS\\
S_{k} & := & \sqrt{diag\left(\Sigma\right)_{k}}\\
\log p\left(S\right) & = & \sum_{k=1}^{K}\log p\left(S_{k}\right)\\
\log p\left(R\right) & = & \log\left(C\left|R\right|^{\eta-1}\right)\\
 & = & \left(\eta-1\right)\log\left|R\right|+C\\
 & = & \textrm{ (LKJ prior)}
\end{eqnarray*}

Let us use a Wishart variational distribution for $\Lambda$:
\begin{eqnarray*}
q\left(\Lambda\right) & = & \textrm{Wishart}\left(V,n\right)\\
E_{q}\left[\Lambda\right] & = & nV\\
E_{q}\left[\log\left|\Lambda\right|\right] & = & \psi_{p}\left(\frac{n}{2}\right)+\log\left|V\right|+K\log2\\
\psi_{p}\left(n\right) & = & \sum_{i=1}^{p}\psi\left(\frac{2n+1-i}{2}\right)
\end{eqnarray*}

Then $\Sigma$ has an inverse Wishart distribution:
\begin{eqnarray*}
E_{q}\left[\Sigma\right] & = & \frac{V^{-1}}{n-K-1}\\
\Sigma_{kk} & \sim & \textrm{InverseWishart}\left(\left(V^{-1}\right)_{kk},n-K+1\right)\\
E_{q}\left[\Sigma_{kk}\right] & = & \frac{\left(V^{-1}\right)_{kk}}{n-K+1-2}=\frac{\left(V^{-1}\right)_{kk}}{n-K-1}\\
\log p\left(\Sigma_{kk}\right) & = & -\left(\frac{\left(n-K+1\right)+1+1}{2}\right)\log\Sigma_{kk}-\frac{1}{2}\frac{\left(V^{-1}\right)_{kk}}{\Sigma_{kk}}+C\\
 & = & \left(-\frac{n-K+1}{2}-1\right)\log\Sigma_{kk}-\frac{\frac{1}{2}\left(V^{-1}\right)_{kk}}{\Sigma_{kk}}+C\\
 & = & \log\left(\textrm{InvGamma}\left(\frac{n-K+1}{2},\frac{1}{2}\left(V^{-1}\right)_{kk}\right)\right)\Rightarrow\\
E_{q}\left[\log\Sigma_{kk}\right] & = & \log\left(\frac{1}{2}\left(V^{-1}\right)_{kk}\right)-\psi\left(\frac{n-K+1}{2}\right)
\end{eqnarray*}

We'll also need the expectation of the square root of an inverse gamma
distributed variable.
\begin{eqnarray*}
p\left(x\right) & = & \frac{\beta^{\alpha}}{\Gamma\left(\alpha\right)}x^{-\alpha-1}\exp\left(\frac{-\beta}{x}\right)\\
E\left[x^{\frac{1}{2}}\right] & = & \int\frac{\beta^{\alpha}}{\Gamma\left(\alpha\right)}x^{-\alpha-1+\frac{1}{2}}\exp\left(\frac{-\beta}{x}\right)dx\\
 & = & \int\frac{\beta^{\alpha}}{\Gamma\left(\alpha\right)}\frac{\beta^{\alpha-\frac{1}{2}}}{\Gamma\left(\alpha-\frac{1}{2}\right)}\frac{\Gamma\left(\alpha-\frac{1}{2}\right)}{\beta^{\alpha-\frac{1}{2}}}x^{-\left(\alpha-\frac{1}{2}\right)-1}\exp\left(\frac{-\beta}{x}\right)dx\\
 & = & \frac{\beta^{\alpha}}{\Gamma\left(\alpha\right)}\frac{\Gamma\left(\alpha-\frac{1}{2}\right)}{\beta^{\alpha-\frac{1}{2}}}\\
 & = & \beta^{\frac{1}{2}}\frac{\Gamma\left(\alpha-\frac{1}{2}\right)}{\Gamma\left(\alpha\right)}
\end{eqnarray*}

Thus
\begin{eqnarray*}
E_{q}\left[\sqrt{\Sigma_{kk}}\right]=E_{q}\left[S_{k}\right] & = & \sqrt{\frac{1}{2}\left(V^{-1}\right)_{kk}}\frac{\Gamma\left(\frac{n-K}{2}\right)}{\Gamma\left(\frac{n-K+1}{2}\right)}
\end{eqnarray*}

This means we have a closed form expectation of the LKJ prior. For
the scale parameters, we can use a gamma prior distribution:
\begin{eqnarray*}
\log p\left(S_{k}\right) & = & \log\Gamma\left(\alpha,\beta\right)\\
 & = & -\beta S_{k}+\left(\alpha-1\right)\log S_{k}+C\\
 & = & -\beta S_{k}+\frac{\left(\alpha-1\right)}{2}\log S_{k}^{2}+C
\end{eqnarray*}

Finally, these expectations are given in terms of the natural parameters, but
for LRVB we need derivatives with respect to the mean parameters. In the Wishart
distribution, the mapping from mean parameters to natural parameters does not
have a closed form.  \eq{basic_lrvb} requires the derivatives of the likelihood
with respect to the moment parameters, and the Hessian must be transformed
before use.  Note that the Hessian of the likelihood is not necessarily at a
maximum, so the transform requires a third-order tensor product.

\end{document}